\documentclass[fleqn,twoside,epsfig]{article}
\usepackage{espcrc2}
\usepackage{amssymb}

\usepackage{graphicx}
\usepackage[figuresright]{rotating}

\newcommand{\be}{\begin{equation}}
\newcommand{\ee}{\end{equation}}

\newcommand{\slh}{\!\!\!\slash}

\title{Towards the Continuum Limit of the Overlap Quark Propagator in Landau Gauge
\thanks{Presented by J. B. Zhang at Lattice 2002}
\thanks{This work is supported by the Australia Research Council}
}
\author{J.\ B.\ Zhang\address[CSSM]{CSSM Lattice Collaboration,\\
Special Research Center for the Subatomic Structure of
Matter (CSSM) and Department of Physics and Mathematical Physics,
University of Adelaide 5005, Australia},
F.\ D.\ R.\ Bonnet\addressmark[CSSM],
P.\ O.\ Bowman\address{Department of Physics and School for Computational Science
and Information Technology, Florida State University, Tallahasse FL 32306, USA},
D.\ B.\ Leinweber\addressmark[CSSM],
A.\ G.\ Williams\addressmark[CSSM]}
\date{\today}

\begin{document}

\begin{abstract}

The properties of the momentum space quark propagator in Landau gauge 
are examined for the overlap quark action in quenched lattice QCD.  
Numerical calculations were done on two lattices with different lattice
spacing $a$ and similar physical volumes to explore the quark propagator
in the continuum limit.  We have calculated the nonperturbative  
wavefunction renormalization function $Z(p)$ and the nonperturbative mass 
function $M(p)$ for a variety of bare quark masses and perform a simple 
linear extrapolation to the chiral limit. We find the behaviour 
of $Z(p)$ and $M(p)$ in the chiral limit are in good agreement between 
the two lattices. 

\end{abstract}

\maketitle
\input epsf

\section{Introduction}

The quark propagator, is one of the fundamental 
quantities in QCD. By studying the
momentum-dependent quark-mass function in the infrared region we can gain valuable insights
into the mechanism of dynamical chiral symmetry breaking and the associated dynamical
generation of mass. The ultraviolet behaviour of the propagator at large momentum
can be used to extract the running quark mass . 

  There have been several studies of the momentum space quark propagator~\cite{jon1,jon2,bowman01,blum01,overlgp}
using different gauge fixing and fermion actions. Here we focus on Landau gauge fixing and
the overlap-fermion action and extend previous work~\cite{overlgp} to two lattices with different lattice
spacings $a$ and very similar physical volumes. This allows us to probe the continuum limit of the 
quark propagator. 

\section{Quark Propagator on the Lattice}
\label{lattice}

In a covariant gauge in the continuum the renormalized Euclidean-space
quark propagator must have the form
\begin{eqnarray}
S(\zeta;p)=\frac{1}{i {p \slh} A(\zeta;p^2)+B(\zeta;p^2)}
=\frac{Z(\zeta;p^2)}{i{p\slh}+M(p^2)}\, ,
\label{ren_prop}
\end{eqnarray}
where $\zeta$ is the renormalization point. 

On the lattice 
the inverse lattice bare quark propagator takes the general form
\be
(S^{\rm bare})^{-1}(p)\equiv
{i\left(\sum_{\mu}C_{\mu}(p)\gamma_{\mu}\right)+B(p)}.
\label{invquargen}
\ee
We use periodic boundary conditions in the spatial directions
and anti-periodic in the time direction.
The discrete momentum values for a
lattice of size $N^{3}_{i}\times{N_{t}}$, with $n_i=1,..,N_i$ and $n_t=1,..,N_t$, are
{\small
\begin{eqnarray}
p_i=\frac{2\pi}{N_{i}a}\left(n_i-\frac{N_i}{2}\right),~~p_t
=\frac{2\pi}{N_{t}a}\left(N_t-\frac{1}{2}-\frac{N_t}{2}\right).
\label{dismomt}
\end{eqnarray}
}
Defining the bare lattice quark propagator as
\begin{equation}
S^{\rm bare}(p)\equiv
{-i\left(\sum_{\mu}{\cal{C}}_{\mu}(p)\gamma_{\mu}\right)+{\cal{B}}(p)}\, ,
\end{equation}
we perform a spinor and color trace to identify
{\small
\begin{eqnarray}
{\cal{C}}_{\mu}(p)=\frac{i}{12}{\rm tr}[\gamma_\mu{S^{\rm bare}(p)}]
,~~{\cal{B}}(p)=\frac{1}{12}{\rm tr}[S^{\rm bare}(p)] \, .
\label{curlyCandB}
\end{eqnarray}
}
The $C_{\mu}(p)$ and $B(p)$ in Eq.(\ref{invquargen}) can be written
\begin{equation}
{\small 
C_{\mu}(p) = \frac{{\cal{C}}_{\mu}(p)}{{\cal{C}}^{2}(p)+{\cal{B}}^{2}(p)}
,~~B(p) = \frac{{\cal{B}}(p)}{{\cal{C}}^{2}(p)+{\cal{B}}^{2}(p)} \, ,
\label{cmup_bp}}
\end{equation}
where ${\cal{C}}^{2}(p)=\sum_{\mu}({\cal{C}}_{\mu}(p))^{2}$. 

 We can identify the appropriate kinematic lattice
momentum $q$ directly from the definition of the tree-level quark propagator,
\begin{eqnarray}
q_\mu\equiv C_{\mu}^{(0)}(p)&=&\frac{{\cal{C}}_{\mu}^{(0)}(p)}{({\cal{C}}^{(0)}(p))^{2}+({\cal{B}}^{(0)}(p))^{2}} \, .
\label{latmomt}
\end{eqnarray}

Having identified
the lattice momentum $q$, we can now
define the bare lattice propagator as
\begin{eqnarray}
S^{\rm bare}(p)
&\equiv & \frac{1}{i{q\slh}A(p)+B(p)}
=\frac{Z(p)}{i{q\slh}+M(p)} \nonumber \\
& = & Z_2(\zeta;a) S(\zeta;p) \, ,
\end{eqnarray}
where $S(\zeta;p)$ is 
the lattice version of the renormalized propagator in
Eq.~(\ref{ren_prop}).

The overlap fermion formalism~\cite{neuberger0,neuberger2}
realizes an exact chiral
symmetry on the lattice and is automatically ${\cal O}(a)$ improved. 
The massless coordinate-space overlap-Dirac operator can be written in
dimensionless lattice units as
\begin{equation}
   D(0) = \frac{1}{2}\left[1 + \gamma_5 \epsilon(H_w)\right] \, ,
\label{D0_definition}
\end{equation}
where $\epsilon(H_w)$ is the matrix sign function,
where
$H_w(x,y)=\gamma_5 D_w(x,y)$ is the Hermitian Wilson-Dirac
operator and where $D_w$ is the usual Wilson-Dirac operator
on the lattice.  However, in the overlap formalism the Wilson mass
parameter $m_w$ enters with a negative sign. 

The massless overlap quark propagator is given by
\begin{equation}
S^{\rm bare}(0) = \frac{1}{2m_w}\left[ D^{-1}(0)-1\right]
                = \frac{1}{2m_w}\tilde D^{-1}(0) \, .
\label{massless_quark}
\end{equation}
This definition of the massless overlap quark propagator follows
from the overlap formalism~\cite{Narayanan:1994gw}
and ensures that the massless quark propagator anticommutes with
$\gamma_5$, i.e.,
$\{\gamma_5,S^{\rm bare}(0)\}=0$ just as it does in the continuum
\cite{edwards2}.  The tree-level momentum-space massless quark
propagator defines the kinematic lattice momentum $q$,
\begin{equation}
S^{\rm bare}(0,p)\equiv \tilde D_c^{-1}(0,p)
       \to S^{(0)}(0,p)=\frac{1}{iq\slh} \, ,
\end{equation}
We can obtain $q$ numerically and analytically from the tree-level massless
quark propagator~\cite{overlgp}. 

Having identified the massless quark propagator in
Eq.~(\ref{massless_quark}), we can construct the massive overlap
quark propagator by simply adding a bare mass to its inverse, i.e.,
\begin{equation}
(S^{\rm bare})^{-1}(m^0)\equiv (S^{\rm bare})^{-1}(0)+m^0 \, .
\label{massive_inverse_prop}
\end{equation}

\section{Numerical results}
\label{numerical}

Here we work on two lattices with different lattice spacing $a$ and
very similar physical volumes  using a tadpole-improved plaquette plus rectangle
gauge action. For each lattice size, we use 50 configurations .
Lattice parameters are summarized in Table~\ref{simultab}. 

\begin{table}
\caption{Lattice parameters.}
\begin{tabular}{ccccc}
\hline
Action &Volume & $\beta$ &$a$ (fm) &  $u_0$\\
\hline
Improved       & $12^3\times{24}$ & 4.60 & 0.125  & 0.88888 \\
Improved       & $8^3\times{16}$  & 4.286& 0.194  & 0.87209 \\
\hline
\end{tabular}
\label{simultab}
\vspace*{-0.5cm}
\end{table}

Our calculations use $\kappa=0.19$  for lattice 1 ($12^3\times{24}$) and $\kappa=0.1864$ 
for lattice 2 ($8^3\times{16}$) 
to make $m_w a= 1.661$ on both lattices. 
We calculate at ten bare quark masses 
in physical units of
$m^0=$ $126$, $147$, $168$, $210$, $252$, $315$,
$420$, $524$, $629$, and $734$~MeV respectively. 

\begin{figure}[tb]
{\small
\resizebox*{\columnwidth}{4.5cm}{\rotatebox{90}{\includegraphics{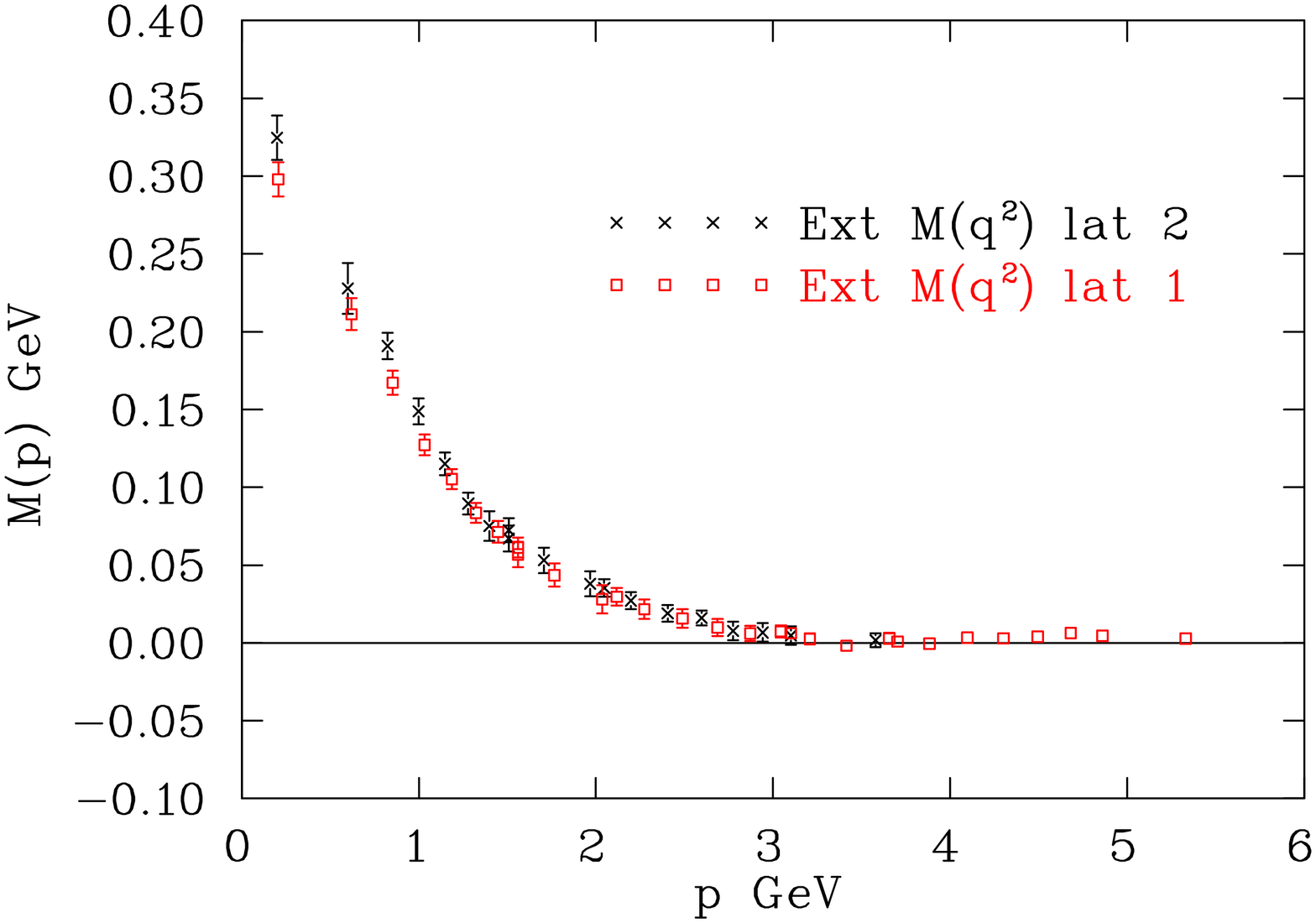}}}
\vspace*{-1.0cm}
\resizebox*{\columnwidth}{4.5cm}{\rotatebox{90}{\includegraphics{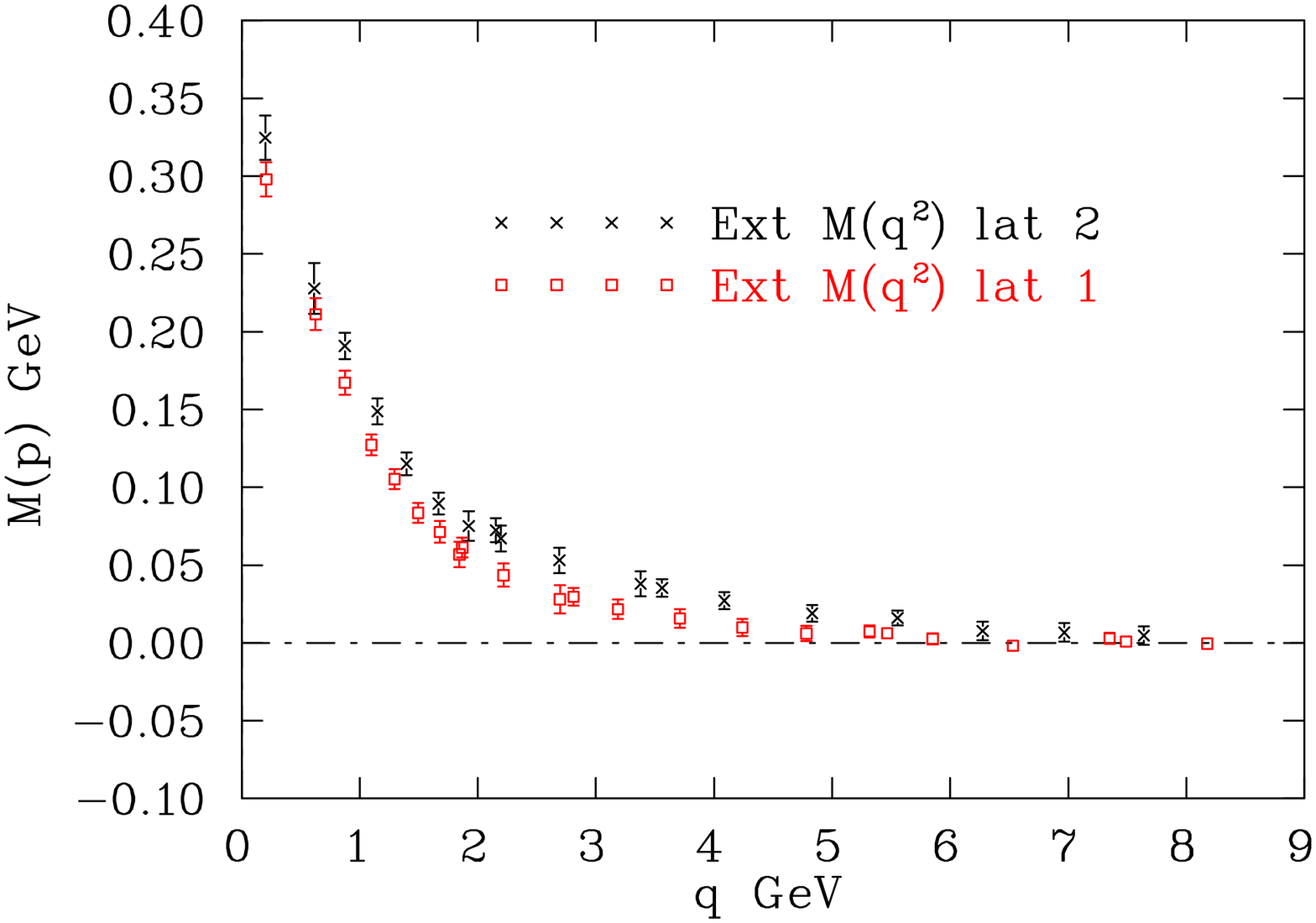}}}
}
\vspace*{-0.5cm}
\caption{\label{ovrmp}{\small Comparison of the mass function $M(p)$ of two lattices in the chiral limit.
The upper graph is plotted against the discrete lattice
momentum $p$ and the lower graph is plotted against the kinematical lattice momentum $q$.}}
\vspace*{-0.6cm}
\end{figure}

\begin{figure}[tb]
\resizebox*{\columnwidth}{4.5cm}{\rotatebox{90}{\includegraphics{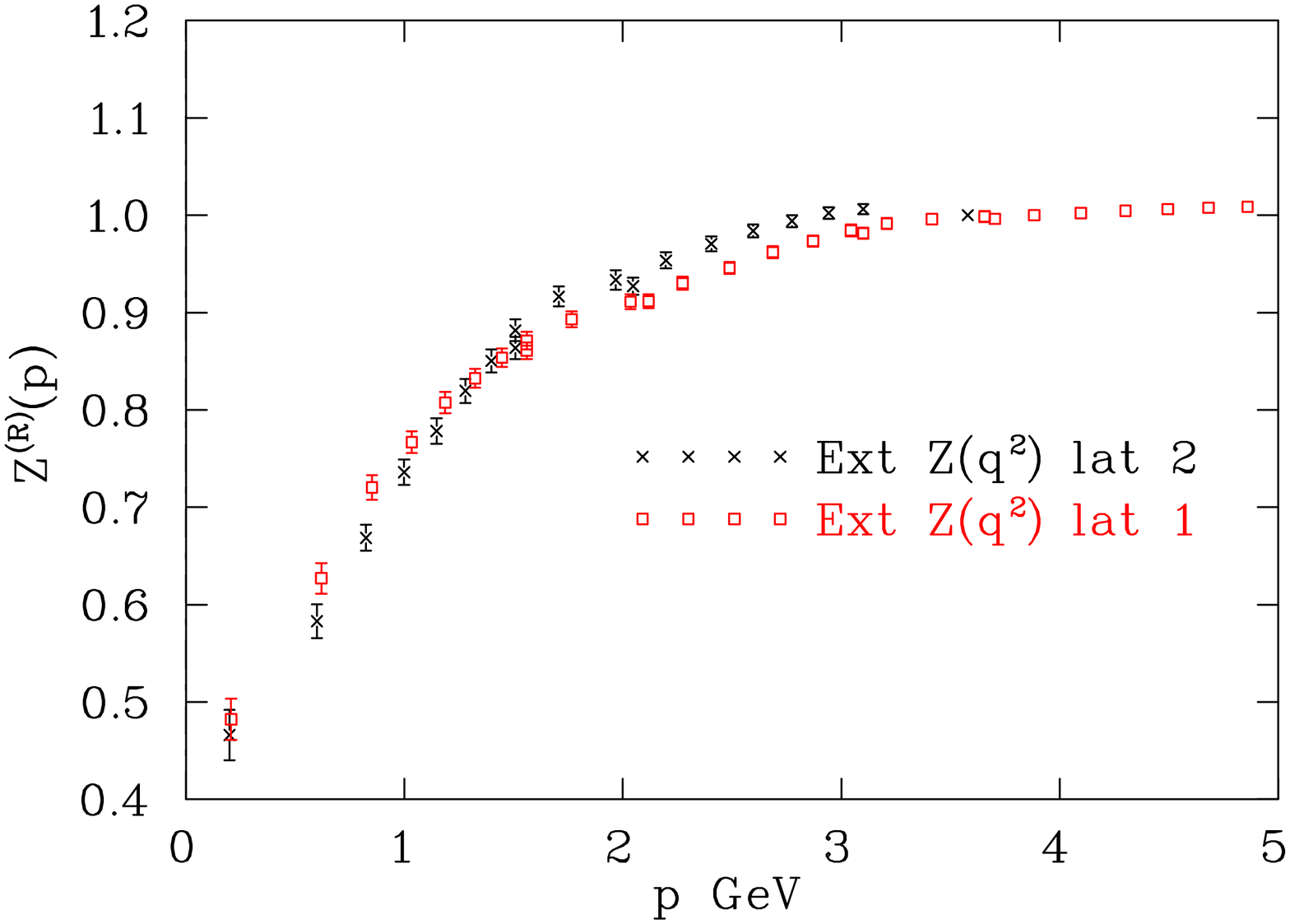}}}
\vspace*{-1.0cm}
\resizebox*{\columnwidth}{4.5cm}{\rotatebox{90}{\includegraphics{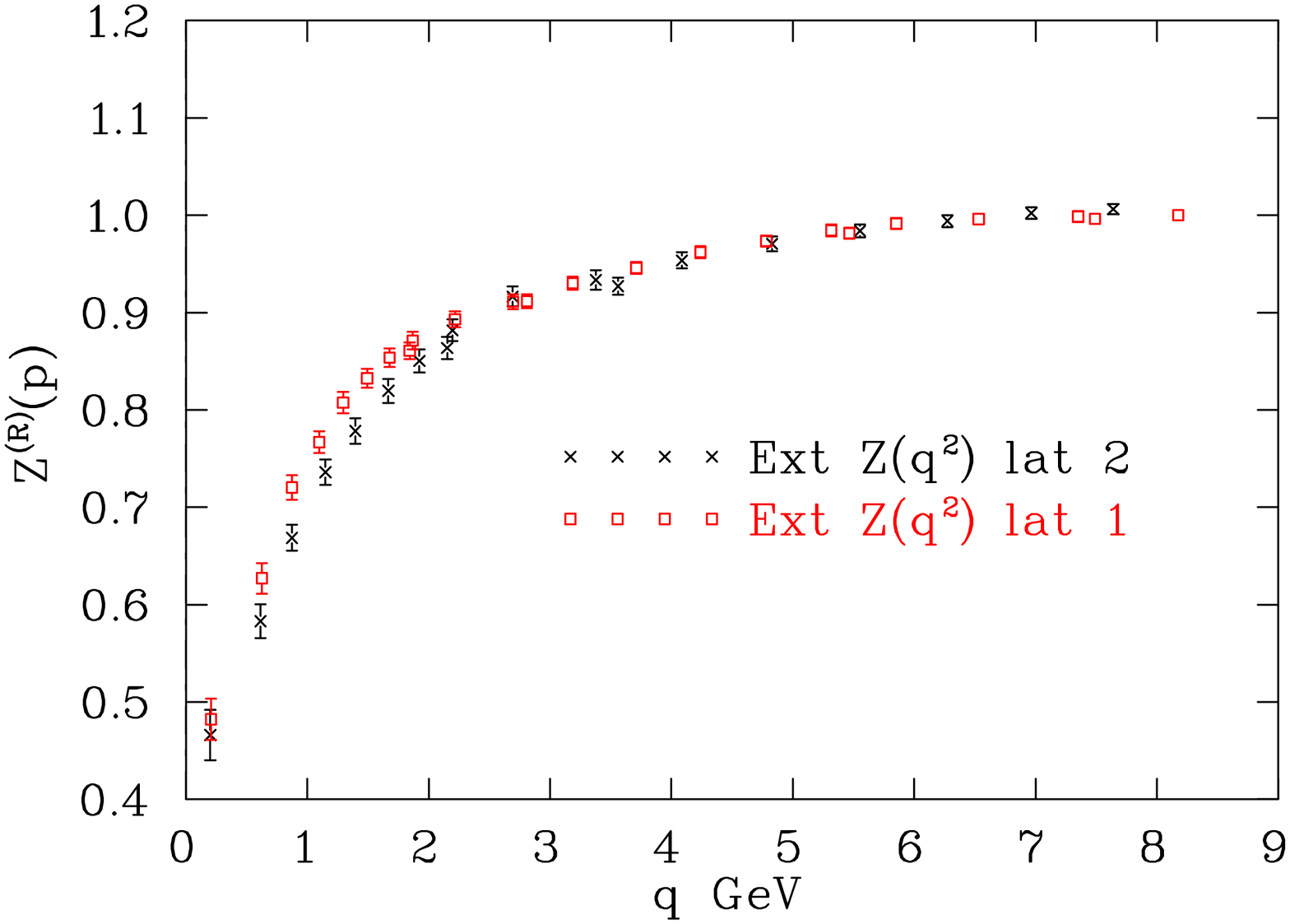}}}
\vspace*{-0.5cm}
\caption{\label{ovrzp}{\small Comparison the renormalization function $Z^{(\rm{R})}(p) $
of two lattices in the chiral limit. 
The upper graph is plotted against the discrete lattice
momentum $p$ and the lower graph is plotted against the kinematical lattice momentum $q$.}}
\vspace*{-0.6cm}
\end{figure}

 The results of lattice 1 are presented in detail in Ref.~\cite{overlgp}. It is
satisfying that the results of lattice 2 are
similar to those of lattice 1. Here we focus on the  
comparision of the results on these two lattices.  All data has been cylinder cut~\cite{overlgp}
and extrapolated to the chiral limit using a simple linear chiral extrapolation . 
The  mass function $M(p)$ for the  two lattices is plotted 
in Fig.~\ref{ovrmp} and the renormalization function $Z(p)$ of the two lattices is plotted in Fig.~\ref{ovrzp}.
We can see that when the mass function $M(p)$ is plotted against the discrete lattice momentum $p$
the results of the two lattices are in good agreement,  
while for the renormalization function $Z(p)$, good agreement is reached on the two lattices if it 
is plotted against the kinematical lattice 
momentum $q$.

\section{Summary}

 In this report, we use tadpole-improved quenched lattice configurations, and  
the overlap fermion operator with the Wilson overlap kernel.
The momentum space quark propagator is calculated in Landau gauge  
on two lattices with different lattice
spacing $a$ and very similar physical volumes to explore 
the continuum limit. 
We calculate the nonperturbative momentum-dependent
wavefunction renormalization function $Z(p)$ and the nonperturbative mass
function $M(p)$ for a variety of bare quark masses and perform a simple
linear extrapolation to the chiral limit. 
We have seen that, the continuum limit for, $Z(p)$, is most rapidly 
approached when it is plotted against the kinematical lattice momentum $q$, whereas
for the quark mass function, $M(p)$, we should plot against the discrete lattice
momentum $p$. The agreement between the two lattices suggests that we are close to 
the continuum limit.



\begin{thebibliography}{10}


\bibitem{jon1}
J.\ I.\ Skullerud and A.\ G.\ Williams,
Phys.\ Rev.\ D {\bf 63}, 054508 (2001) ;
Nucl.\ Phys.\ Proc.\ Suppl.\  {\bf 83}, 209 (2000) .

\bibitem{jon2}
J.\ Skullerud, D.\ B.\ Leinweber and A.\ G.\ Williams,
Phys.\ Rev.\ D {\bf 64}, 074508 (2001) .

\bibitem{bowman01}
P.\ O.\ Bowman, U.\ M.\ Heller and A.\ G.\ Williams,
Nucl.\ Phys.\ B (Proc.\ Suppl.) {\bf 106}, 820 (2002) .

\bibitem{blum01}
T.~Blum {\it et al.},
Phys. Rev. D{\bf 66},014504(2002).

\bibitem{overlgp}
F.\ D.\ R.\ Bonnet, P.\ O.\ Bowman, D.\ B.\ Leinweber,
A.\ G.\ Williams and J.\ B.\ Zhang,  Phys.\ Rev.\ D {\bf 65}, 114503 (2002) .

\bibitem{neuberger0}R.\ Narayanan and H.\ Neuberger {Nucl. Phys.} {\bf B443},
305 (1995) .

\bibitem{neuberger2}H.\ Neuberger, {Phys. Lett.} {\bf B427}, 353(1998).

\bibitem{Narayanan:1994gw}
R.\ Narayanan and H.\ Neuberger,
Nucl.\ Phys.\ B {\bf 443}, 305 (1995) .

\bibitem{edwards2} R.G. Edwards, U.M. Heller, and
R. Narayanan, {Phys. Rev. D} {\bf 59}, 094510(1999).

\end{thebibliography}
\end{document}